# Archaeo-Astronomy in Society: Supporting Citizenship in Schools Across Europe

Daniel Brown, Nottingham Trent University SST/CELS, UK
Lina Canas, Centro Multimeios de Espinho, Navegar Foundation, Portugal

*Abstract: The interdisciplinary topic of archaeo-astronomy links science subjects such as astronomy with archaeology and sociology to explore how ancient societies perceived the heavens above. This is achieved by analysing ancient sites such as megalithic monuments (e.g. Stonehenge), since they are the most common remains of these societies and are wide spread in Europe*
*We discuss how archaeo-astronomy and ancient sites can be transversal to many topics in school. The links to the science curricula in different countries are highlighted. However, especially the subject of citizenship can be supported by exploring the diversity of culture, ideas, and identities including the changing nature of society in the past millennia.*
*We conclude that archaeo-astronomy offers many opportunities for citizenship. Learning more about megalithic monuments in different countries (e.g. England, Portugal, and Germany) supports tolerance and understanding. Furthermore, the distribution of these sites lends itself to explore beyond borders, introduce international networking, and truly develop students into global citizens.*

## 1 Introduction

Archaeo-Astronomy is a highly interdisciplinary subject. The name itself describes it as a collaboration of astronomy and archaeology to explore how mankind began to perceive, visualise and understand the sky above, especially the Sun and the Moon, as early as prehistoric times. This also includes how he used the movement of these celestial bodies to establish e.g. calendars. But archaeo-astronomy goes even further than that by including anthropology and ethno-history (many examples can be found in Ruggles and Saunder, 1993). Gaining insights into the beliefs of prehistoric mankind is challenging, since there are no written records and only their ancient monuments can be used. Famous European sites of interest for archaeo-astronomy are Stonehenge (United Kingdom), the circular ditches at Goseck (Germany), standing stones at Carnac (France), or the Almendres stone circle (Portugal).

Recently there has been a more detailed global survey of such sites by Ruggles and Cotte (2010). This collaborative project between the International Council on Monuments and Sites (ICOMOS), an advisory body to UNESCO for cultural heritage, and the International Astronomical Union (IAU), created a thematic study of Heritage Sites of Astronomy and archaeo-astronomy in the context of the UNESCO World Heritage Convention. They carried out a systematic surveys of over 3000 tombs and 'temples' in Europe as potential heritage sites relating to astronomy and archaeo-astronomy that might have the potential to demonstrate outstanding universal value. As a result they listed some of the worlds most important sites related to astronomical heritage including such sites related to prehistoric Europe, e.g. megalithic stone circles. This work further underlines the wide spread nature of such sites illustrated in figure 1, ranging from Portugal to Poland and from the northern parts of the United Kingdom to southern Italy. The following examples illustrate how close ancient monuments are to highly populated areas allowing them to be freely explored. The United Kingdom offers a wealth of sites beyond Stonehenge, including the stone circles within the National Peak Park District (e.g. Arbor Low) 20 km away from Manchester or Sheffield. In Portugal there is a high density of megalithic tombs (seven-stone antas) located close to the towns of Évora and Elvas. In Germany, even though not explicitly mentioned in the report, there are many sites that can be put into an archaeo-astronomy context. An interesting example being the now destroyed circular ditch structure at Bochum-Harpen in the centre of the highly populated Ruhr area and in proximity to an astronomical theme park containing a so called horizon observatory inspired by Stonehenge and developed by 'Horizontastronomie im Ruhrgebiet e.V.' (Ini Kreis).

This accessibility has supported the imagination in the general public and generated a lively discussion since the 1960. As a result this subject has had a history of many highly speculative theories not always founded on thorough research. Only in the past two decades has it outgrown these limitations and been set on a very solid scientific basis (Ruggles, 1999). During this time ancient monuments such as stone circles have been discovered as an ideal environment to spark debate and creativity, see e.g. the Gardom's Edge project (Bevan *et al.*, 2004).

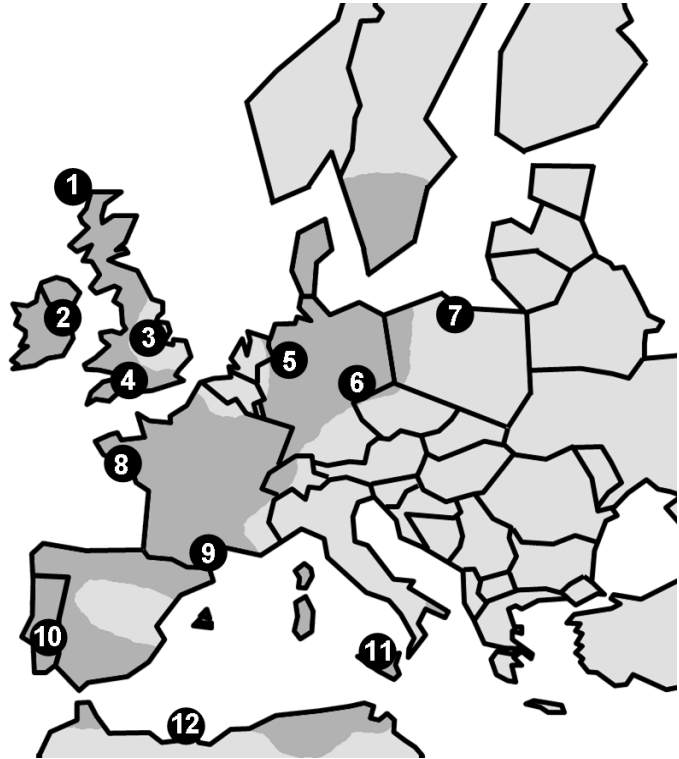

**Figure 1** – The distribution of ancient megalithic monuments throughout Europe. The shaded areas illustrate the spread of ancient megalithic minumets and the labels point out specific megalithic monuments: 1-Callanish, 2-Newgrange, 3-Arbor Low, 4-Stonehenge, 5-Bochum-Harpen, 6-Goseck, 7-Odry, 8-Carnac, 9-Morrel das Fadas, 10-Almendres, 11-Mura Pregne, and 12-Beni Messous.

The layout of megalithic monuments can range from mere circular ditches dug into the ground with openings, passage tombs (e.g. Newgrange), or impressive rings of standing stones surrounded by banks and ditches. Many of them include a combination of simple geometrical shapes or alignments into their design. These features indicate in some specific cases rising or setting points of the Sun or the Moon on the horizon. The monuments are also closely embedded into their surrounding landscape. This connection is established not only by using local resources and structuring the landscape they are placed in, but also by using the profile of the landscape to highlight intended alignments.

These sites can act as laboratories exploring aspects of mathematics, sciences, history, sociology, and even religion in an educational context, given their prehistoric context, architectural set-up, and clear integration into the landscape. Their educational potential can be understood within the recent debate on what science is supposed to achieve in schools that focused on the outcome of Scientific Literacy. A possible definition of such literacy has been given by Durant (1993) and includes three aspects: scientific facts, scientific method and experiments, and science in the scientific community. The student should explore the scientific method and its different applications by carrying out experiments themselves. Although this can be done within the classroom environment, a review of recent studies on outdoor learning and teaching by Rickinson *et al.* (2004) has shown that the outdoor-classroom furthers deeper learning and demonstrates the practicality of taught skills. Ancient historic monuments such as megalithic stone circles offer a wealth of opportunities to explore with schools and such work has been supported by e.g. the archaeo-astronomy project in the UK (Brown, Francis, and Alder 2010).

In the following sections we will describe in more detail how archaeo-astronomy fits into the school curriculum for the three examples of England, Germany, and Portugal. The most important links to citizenship are mentioned and their potential to further the teaching of citizenship are discussed in section 3. We then summarise our findings and draw some final conclusions in the final section of our paper.

## 2 Archaeo-Astronomy in Schools Across Europe

Given the inherent multi-disciplinary nature of archaeo-astronomies character there are many links with schools curricula that have been outlined in Brown, Neale and Francis (2010) and will be briefly summarised. As a topic archaeo-astronomy can be introduced into schools to explore certain aspects of astronomy taught in the physics curriculum. The concept of how the movement of the Sun during a day and a year moves can be revisited. This will lead inevitably onto a discussion of how seasons are created. A possible usage of ancient sites to observe the celestial bodies can be analysed after they have been surveyed in a certain manner. This survey involves applying numeracy skills and the ability to make justified approximations. During this process the ancient monument has been used as a mathematical laboratory outside of the classroom. Pupils can experience for themselves how their more abstract knowledge of numbers and geometry applies to the real life within the real world. During this exploration of an ancient site the question will be raised: How were they built and why? To address these questions the subject of archaeology or history will be targeted. Principles such as the three age system can be introduced or how the stratification of a site illustrates the sequence of the construction. But some aspects of the function of the site will be hard to imagine, since the surrounding landscape will have changed during the centuries. Environmental aspects of biology can now be introduced to explore how mankind changed the landscape surrounding him.

Overall, there is a general theme apparent in the application of archaeo-astronomy. When addressing each subject the pupils become aware of their place within the landscape, society, and in history. They will have developed their empathy skills and critical thinking. These skills lead to revisiting many misconceptions e.g. regarding the intelligence of ancient cultures. Therefore, working with archaeo-astronomy strongly supports citizenship. The importance of including science in the society to support scientific literacy in science lessons has been mentioned by Durant (1993). It is essential for students to learn that science has been developed by humans living within a society. Given different constraints the approaches, concepts, and resources will be different.

We will explore in more detail how the above mentioned aspects apply to schools in England, Germany, and Portugal.

### 2.1 Schools in England

The English educational system consists of two stages. Students enter primary school at an age of five years for six years after which they then proceed into secondary school until the age of 16 or 18. Within the English science curriculum aspects of archaeo-astronomy can be linked to key concepts introduced at Key Stage (KS) 1 and KS2 at primary school, as well as KS3 and KS4 at secondary school.

- The astronomy side can be used to support light and shadows as well as the apparent movement of the Sun taught in KS2 'Earth and beyond' (Sc4 4b-d) in 'Physics processes'. Furthermore, the more detailed analysis of seasonal changes of the path of the Sun will help to consolidate knowledge in 'Environment, Earth, and Universe' at KS 3 (3.4b).
- The archaeological concept of stratigraphy is included into the geological activity caused by chemical and physical processes that is integrated into 'The environment, Earth, and the Universe' (3.4a) at KS3. Later, at KS4, archaeobotany can be used to explore how organisms are interdependent and adapt to their environment within 'Organisms and health program' (2.1a) or the effects of human activity on the environment covered in 'Environment, Earth, and Universe' (2.4a).
- The biological and ecological aspects are covered in 'Life processes and living things' through all ages. At KS1 the topic supports finding out about the animals and plants that live in local habitats (Sc2 5a) and at KS4 it targets how humans impact on the environment depending on social and economic factors (Sc2 5b). Additionally, the surveying and data handling skills are in line with the investigative skills section of the curriculum.

Looking beyond the science curriculum, archaeo-astronomy also offers many links to Citizenship. This subject has been trialled and tested in the UK by different groups during the past decade. Especially its links to science have been investigated by the Science Education Group at the University of York (Burden 2005). As a result from their work, citizenship has since 2006 become a subject within the English and Welsh curriculum. Although schools are not required to teach this subject, it is included when teaching any other subject including science (see e.g. Brown and Neale 2010). Therefore, England and Wales fully embrace the cross-curricular aspect of Citizenship and support scientific literacy.

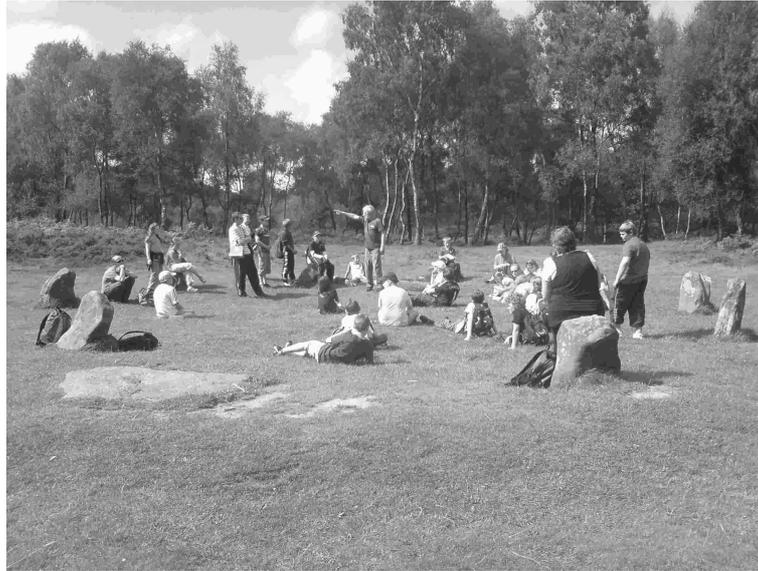

**Figure 2** – Secondary and primary students explore the Nine Ladies stone circle, including dangers of quarrying and recent demonstrations.

On the subject of archaeo-astronomy, students can explore certain examples of stone circles and experience how different groups in the community value these sites. In the case of the Nine Ladies site shown in figure 2, students can be introduced into the problems of quarrying in the peak district and link this to the aspects of citizenship related to the program 'Rights and responsibilities' in KS 3 (1.2b). Furthermore, rock art located at some sites offers a focus to apply empathy to interpret its possible use.

## 2.2 Schools in Germany

The educational system in Germany is split into three cycles. Students with an age of five years enter the *Grundschule* for four years. They can then either: complete and leave after *Sekundarstufe* I in a *Hauptschule* or *Realschule* with an age of 15 years, or complete *Sekundarstufe* I and II in a *Gymnasium* or *Gesamtschule* and leave with an age of 18 years.

Germany is a federalist union, consisting of 16 confederate states, each of which having autonomy in the details of its educational system. Each state has an official curriculum and there are 160 different schoolbooks in use throughout German schools. However, the teacher does not have to strictly follow the contents within a topic. This freedom allows him to teach more effectively in light of many inconsistencies within the curriculum. Therefore it is very difficult to give a detailed list of links to a curriculum. We will only point out links with the curriculum in the state of Northrhine-Westphalia (Ministerium für Schule und Weiterbildung NRW, 2008a, 2008b, 2008c, 2008d, 2008e) as experienced by a student up until *Sekundarstufe* I in a *Gymnasium*.

- Astronomy links are present in the lesson plans of *Sachkunde*. During the *Grundschule*, students experience aspects of Light and Shadow in the topic of 'nature and life' focussing on 'animals, plants, environments'. Furthermore, this topic is extended with aspects of seasons when focussing on 'heat, light, fire, water, air, sound'. In the topic of 'space, environment and mobility' the daily motion of the Sun is used to determine directions when focussing on 'school and surroundings'. When entering *Sekundarstufe* I, astronomy is only present when dealing with energy aspects, e.g. in the contents area of 'temperature and energy' or 'Sun – temperature – Seasons'.
- Archaeological links can be found during the *Grundschule* (*Sachkunde*) in the topic of 'space, environment and mobility' within 'home and world' or in 'time and culture' within 'past and present'. Here the students explore geographical changes in a region and beyond, including relevant links with stratigraphy. Furthermore, the students explore the past history of their region. Later in *Sekundarstufe* I they can extend their knowledge about early cultures in the subject of history. Archaeobotany can be used to further deepen the aspects explored in the subject of biology in 'diversity of creatures and adaptation of plants and animals to the seasons'.

- Environmental aspects are explored throughout the *Grundschule* (*Sachkunde*). The interaction between humans or animals with their environment is covered in the topic of 'nature and life' when focussing on 'animals, plants, environments'. This is covered during *Sekundarstufe* I in the subject of biology in the contents area of 'diversity of creatures'.

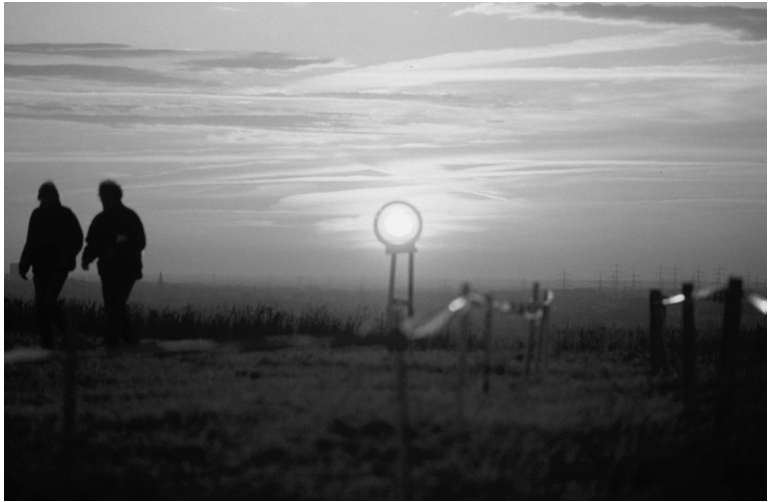

**Figure 3** – German gymnasium students rebuilding Stonehenge and exploring construction methods of ancient monuments. A project initiated by their science teacher M. Winkhaus a member of Ini Kreis who provided the image.

Citizenship as a subject is not included in any curriculum at German schools at any stage. Given the political past in Germany, there is no specific focus upon moral standards or values in e.g. the subject related to political education. For a more in depth discussion see Himmelmann (2004). However teachers are obliged to teach values, but can choose freely the way in which this is achieved. Especially in *Grundschule* there are several possible links offered in *Sachkunde*. Within the topics of 'humans and community' empathy is included and in 'time and culture' living conditions now and in the past as well as differences between cultures, lifestyles and religions are explored. Later in politics *Sekundarstufe* I only the contents area of 'identities and shaping life in the changes of modern society' allows for a more modern inclusion of citizenship. This has lead to citizenship related projects being initiated by teachers themselves on e.g. xenophobia or in one case even archaeo-astronomy by the Ini Kreis as shown in figure 3. But only very recently has there been a discussion to include citizenship aspects into science to create a closer connection between science and the general public (Allgaier 2007). This new trend offers an ideal opportunity to include the above mentioned links with cultures through time into the science curriculum via archaeo-astronomy activities.

### 2.3 Schools in Portugal

In Portugal there are three main cycles of basic education in an overall of nine years of basic mandatory education. The first cycle embraces the first four years, the second cycle the following two years and the third cycle the last three. Besides disciplinary subjects, there are also other areas to be developed within class hours that are integrated on the mandatory curriculum called: Project Area, Guided Study and Civic Education. In the Portuguese curriculum Citizenship Education is also integrated as a general guide line, transversal to all subjects and subject areas whether disciplinary or non-disciplinary (more details can be found in Alves *et al.*, 2003).

The Portuguese science curriculum is focused into four main themes: Earth in space, Earth in transformation, sustainability on Earth and better living on Earth. All these themes involve scientific, technological, social and environmental components:

- *Earth in Space*: In the *1st Cycle* in *Environment Studies* subject and in *3rd Cycle*, in *Physics and Chemistry,* students are encouraged to observe their surroundings, reinforcing concepts such as the motion of Earth through space, the occurrence of seasons by observations of the Sun during the year. This knowledge can be applied by students of all three cycles to the example of the several seven-stone antas (see figure 4) whose principal orientation lies within the arc of sunrise.

- *Earth in transformation*: In all three cycles in *Natural Sciences* subject students observe their surroundings and start using criteria to classify materials existing on Earth. This can be linked to observing and analyzing the building blocks of megalithic sites such as granite that are present in most of seven-stone antas and, in some other cases even quartz schist.
- *Sustainability on Earth* and *Better living on Earth*: During the *1st cycle* students are encouraged to perform surveys of situations in the local environment such as monuments and other constructions. Archaeo-astronomy links continue until the *3rd cycle* where they explore the impact on the deterioration of monuments. Thereby, the application illustrates how human intervention on Earth affects the surrounding environment.

The non-disciplinary Project Area is the perfect example where archaeo-astronomy related activities can be included to their fullest potential. Besides the several links with the science program described above, archaeo-astronomy related activities can embrace other subject areas such as History, Geography or even Artistic Education.

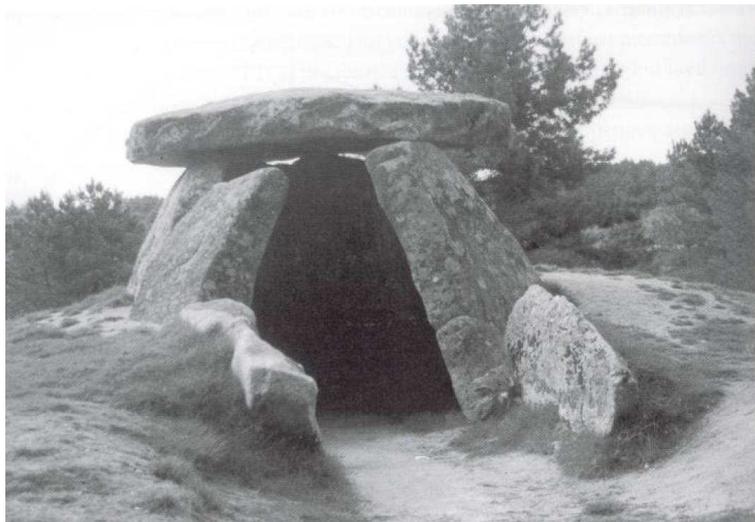

**Figure 4** – Fonte Coberta, a seven-stone anta near Chã de Alijó, Miranda Plateau. This is a typical example of seven-stone antas in Portugal. Image is taken from Hoskin (2001).

### 3 An Opportunity for Citizenship

The possible curriculum links with archaeo-astronomy are diverse and cover a wide range of subjects and age groups across different countries, illustrated in the section above. The most important opportunity utilising the international aspects has been offered by the subject of citizenship.

When exploring archaeo-astronomy, students automatically have to place themselves into the frame of mind of ancient mankind. They will experience how their initial picture of Stone Age brutes wilding clubs becomes replaces by a far more complex and intelligent society. This process fosters the key aspects of citizenship through the development of empathy, as specifically stated in the English / Welsh and German curriculum. It can easily be combined with more modern misconceptions and recent social tension in their country. Furthermore, the analysis of sites within and outside the boundary of their own country opens the mind of the students to different cultural groups and identities: some are not bound to just one country and others can be part of our cultural heritage. Again, this allows dealing with modern topics within citizenship such as xenophobia in an innovative way.

Apart from these social aspects of citizenship, sustainability and environmental responsibility is also part of citizenship. Archaeo-astronomy supports this area by introducing students into how ancient monuments are embedded into and are part of the landscape, thereby capturing certain values and meanings related to this place. These aspects are present in the curriculum of all countries discussed above. Both England and Portugal have megalithic monuments that are or have been under threat of destruction either by environmental pollution or quarrying. In Germany many of the interesting circular ditches within the Ruhr-Area are e.g. buried under motorways. These locations are an ideal focus to discuss responsible behaviour within the environment and dealing with cultural heritage. Students can also learn more about the many different groups claiming interest in such sites. The impact of different non-environmental friendly activities (e.g. quarrying and mining) can have very surprising results. The German astronomy theme park that has been built on top of a now regenerated slag heap illustrates this fact. It contains a horizon observatory inspired by the design of Stonehenge and embraces the local former mining and steel industry (Chadha 2006). It offers the opportunity for the general public to learn more about megalithic monuments, their astronomical background, and how they expresses perceived values of the region.

Social competence and sustainability are the two main factors within citizenship that archaeo-astronomy can support. The wide spread distribution of megalithic monuments throughout Europe is ideal to visualise all the aspects of becoming a global citizen and raise its importance within the curriculum e.g. in Germany. Fostering inter-European collaboration can only be of benefit for archaeo-astronomy and citizenship in general.

## *4 Conclusions*

We have outlined the key elements present in science curricula for some European countries: England, Germany and Portugal. In these countries we identified common key areas targeted by archaeo-astronomy: Planet Earth and its place in the Universe as well as mankind as an active, changing and perceptive member of his surroundings (environmental and social). Especially the topics of alignments, geological composition, historical and social relevance are present in the innovative study of ancient megalithic sites with archaeo-astronomy. We elaborated how specific areas of science curricula across three countries are targeted and gave details:

- **Astronomy:** Light and shadow and seasonal changes of the path of the Sun
- **Geology:** Stratigraphy
- **Biology:** Archaeo-botany
- **Chemistry:** Effects of industrial and technological development

Understanding how ancient mankind has influenced and changed his surrounding environment and that this is still happening at a faster and sometimes worrying and dangerous pace, leads to an equal gain in a responsible attitude towards sustainability and preservation of the environment. Through practical examples such as megalithic site analysis, archaeo-astronomy fully offers a chance for students to cover a wide range of subjects beyond science. These experiences are gained 'close to home' in the outdoor classroom supporting scientific literacy.

Overall, this is a valuable opportunity for Citizenship Education. By studying ancient megalithic sites the students have the opportunity to develop values such as respect for cultural heritage leading to ecological and social awareness. Additionally, they will learn about civic intervention in a creative way and explore their identity.

While the students are exploring all these additional values and meanings embedded in the landscape, they have been experiencing aspects of place-based teaching. This educational approach is based upon the concept of 'sense of place' that can be defines as: "What begins as undifferentiated space becomes place when we endow it with value" (Tuan 1977). Although the idea of sense of place is not new within humanistic subjects, place-based teaching has only been integrated into science in the past decade and is described by Semken and Freeman (2008). It includes aspects of emotional bonds, values, qualities, place meanings, and awareness of cultural heritage described in more detail by Williams and Stewart (1998). After participating in archaeo-astronomy based activities the students have engaged in many of the aspects mentioned, become quite aware and grounded within the region they call home.

Given the wide distribution of megalithic sites across Europe, citizenship can be supported in this manner in many countries beyond the stated examples. A local ancient site can offer the opportunity to share and relate archaeo-astronomy findings as well as values at a European or even global level. Only by discovering ones roots and place of origin can one become global. The collaboration can then gather teachers, students and the general community and reinforce awareness towards historical site preservation and a sense of place, creating truly global citizens.

## About the Authors

*Dr. Daniel Brown*
Daniel Brown is a professional astronomer who graduated in Germany and carried out his PhD in the UK. He is now working at the Nottingham Trent University and its on-site observatory, where he supports astronomy teaching and outreach work with the general public and schools. This also includes working with creative practitioners and theatre groups. The main focus of his outreach work is based on archaeo-astronomy and the use of the outdoor-classroom in schools, further, and higher education. Furthermore, he is a founding member of the 'Horizontastronomie im Ruhrgebiet e.V.', a German private initiative promoting astronomy outreach based on an EU funded Science Park located within the ruhr area.

*Lina Canas*
Lina Canas graduated in Astronomy at the Faculty of Science of the University of Porto, and in 2008 completed a Masters in Geophysics at the same university. She works at the Planetarium and Observatory at Centro Multimeios de Espinho, being part of the team responsible for the production, development and implementation of activities related to science communication. Of her most recent projects, we can highlight the executive production of the planetary show Journey to a Black Hole; Dinner on Mars, a project that won the international first prize for the "Most Innovative Event" during the "Galilean Nights " of the International Year of Astronomy 2009; Camping at the Planetarium which got an international honourable mention for the" Most Innovative Event "during the" 100 Hours of Astronomy " of the International Year of Astronomy 2009. She is also part of the Science Office team.